\documentclass[journal, twocolumn]{IEEEtran}
\usepackage[dvips]{graphicx}\DeclareGraphicsExtensions{.eps}
\usepackage{amssymb}
\usepackage{amsmath}
\usepackage{amsbsy} 
\usepackage[mathcal]{eucal}
\usepackage{subfigure}
\usepackage{cite}
\usepackage{geometry}
\newcommand{\argmax}{\arg\!\max}

\begin{document}

\title{Distributed Beam Scheduling for Multi-RAT Coexistence in mm-Wave 5G Networks}

\author{\IEEEauthorblockN{Maziar Nekovee, Yinan Qi and Yue Wang}\\
\IEEEauthorblockA{Samsung Electronics R\&D Institute UK, Staines, Middlesex TW18 4QE, UK}} 
\maketitle

\begin{abstract}
Millimetre-wave communication (licensed or unlicensed) is envisaged to be an important part of the fifth generation (5G) multi-RAT ecosystem. In this paper, we consider the spectrum bands shared by 5G cellular base stations and some existing networks, such as WiGig. Sharing the same band among such multiple radio access technologies (RATs) is  very challenging due to the lack of centralized coordination and demands novel and efficient interference mitigation and coexistence mechanisms to reduce the mutual interference. To address this important challenge, we propose in this paper a novel multi-RAT coexistence mechanism where neighbouring 5G and WiGig base stations, each serving their own associated UEs, schedule their beam configurations in a distributed manner such that their own utility function, e.g. spectral efficiency, is maximized. We formulate the problem as a combinatorial optimization problem and show via simulations that our proposed distributed algorithms yield a comparable spectral efficiency for the entire networks as that using an exhaustive search, which requires global coordination among coexisting RATs and also has a much higher algorithmic complexity.

\end{abstract}


\section{Introduction}\label{sec:intro}
One of the primary contributors to global mobile traffic growth is the increasing number of wireless devices that are accessing mobile networks. Over half a billion (526 million) mobile devices and connections were added in 2013 and the overall mobile data traffic is expected to grow to 15.9 exabytes per month by 2018, nearly an 11-fold increase over 2013~\cite{Cis14}. 

In order to address this issue, a recent trend in 3GPP is to utilize both the licensed and unlicensed spectrum simultaneously for extending available system bandwidth. In this context, LTE in unlicensed spectrum, referred to as LTE-U, is proposed to enable mobile operators to offload data traffic onto unlicensed frequencies more efficiently and effectively, and provides high performance and seamless user experience~\cite{RML+15}. Integration of unlicensed bands is also considered as one of the key enablers for 5G cellular systems~\cite{Sam15}. However, unlike the typical operation in licensed bands, where operating base stations (BS) have exclusive access to spectrum and therefore are able to coordinate by exchanging of signaling to mitigate mutual interference, such a multi-standard and multi-operator spectrum sharing scenario imposes significant challenges on coexistence in terms of interference mitigation. 

\begin{figure} [t] 
\centering
\includegraphics[width=6cm, angle =0]{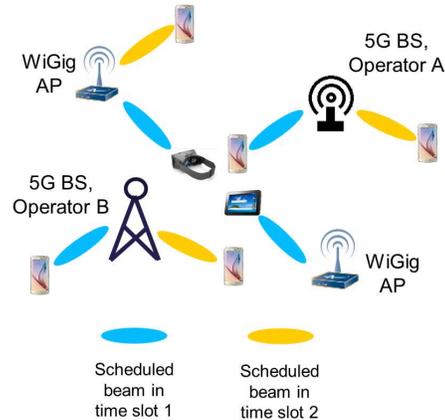}
\caption{Coexistence scenario in 60 GHz deployment with beamforming}
\label{figworstcase}
\end{figure}

Licensed Assisted Access (LAA) with listen-before-talk (LBT) protocol has been proposed for the current coexistence mechanism of LTE-U~\cite{Jeo15}. For coexistence in 5G, one of the major issues is that the use of highly directional antennas as one of the key enablers for 5G networks \cite{TSR13,Maz15} becomes problematic for the current coexistence mechanisms where omni-directional antennas were mostly assumed. For example, transmission by a different nearby 5G BS or WiGig access Point (AP) may not be detected due to the narrow beam that has been used, resulting a transmission that causes excessive interference to the victim user equipment (UE), e.g., UEs in the central area as illustrated in Fig.~\ref{figworstcase}. In this regard, the simultaneous transmission should be coordinated and exploited fully to greatly enhance the network capacity. Such a mechanism is referred as beam scheduling. There has been some related work on this topic based on TDMA \cite{AH08,PKW+09} and a concept of exclusive region is introduced in \cite{CSM10} to enable concurrent transmission with significant interference reduction. However, the effect of interference aggregation is not captured. Other approaches are proposed based on centralized coordination in \cite{GSA+10,SMM10}, where the access points are coordinated in a centralized manner to reduce interferences and improve network capacity. 

In this paper, we consider a multi-RAT deployment where 5G BSs and other 60GHz APs, e.g., WiGig APs, co-exist, all transmitting via beamforming. We form a scheduled beam sequence containing the indices of the beams used at different time slots, and formulate an optimization problem to find the optimal scheduled beam sequence to maximize the spectral efficiency of the entire network. It is known that such a combinatorial problem is NP-hard and highly computationally costly when using exhaustive search \cite{LT15}. We therefore further propose a novel distributed learning algorithm where different BSs cooperatively and iteratively update the beam sequences such that near maximum spectral efficiency is achieved. It is shown that the proposed algorithm almost achieves comparable spectral efficiency to that using the exhaustive search, while at the same time having a much reduced complexity and signaling overhead. It should be noted that the proposed algorithm and analysis are general and can be easily applied to other bands, such as 28 GHz. The rest of the paper is organized as follows. System model will be introduced in the next section and the optimization problem is formulated in section III. In section IV, we detail the proposed distributed learning algorithm and compare its complexity with that using exhaustive search and distributed greedy scheduling. Simulation results are presented in section V and Section VI concludes the paper.

\section{System Model}\label{sysmodel}
In this paper, we assume that both 5G BSs and co-existing APs employ beamforming to tackle the increased path loss in 60 GHz band. From now on, we do not differentiate 5G BS and co-existing AP and refer to them as 5G AP for simplicity. We also assume that each 5G AP is only able to transmit data using a single beam at a time for simplicity and the mechanism considered here can be extended to the multi-beam case. 

We assume a coexistence deployment scenario with $N$ 5G APs, which could either be 5G BSs or co-existing APs or a mixture of both, and $M$ associated UEs for every AP. Each 5G AP has $N_t$ transmit antennas, whilst each UE has one receive antenna. At a given time, the $n$th ($n=1,\dots, N$) 5G AP transmits to the $m$th ($m=1,\cdots, M$) UE using a beam $\mathbf{w}_{nm}$, where $\mathbf{w}_{nm}$ is vector with length $N_t$. To obtain the beamforming vector $\mathbf{w}_{nm}$, we assume that the 5G AP selects the beam configuration within a predefined beam codebook with cardinality $C$ that uniformly covers the azimuth directions around the AP. In particular, the codebooks at the transmitter are formed by vectors $\{\mathbf{v}_{1}, \cdots, \mathbf{v}_{C}\}$, with the $i$th length-$N_t$ vector $\mathbf{v}_i$ denoting the beam for the $i$th codebook entry. The $n$th AP selects the $\hat{i}$th entry in the codebook according to
\begin{equation}
\hat{i} = \argmax_{i=1,\cdots, C}\left|\mathbf{v}_i^T\mathbf{h}_{mn}\right|^2
\end{equation}
and
\begin{equation}
\mathbf{w}_{nm} = \mathbf{v}_{\hat{i}}.
\end{equation}

We define a scheduling cycle with duration of $M$ time slots. Within each scheduling cycle, we consider scheduling the beams for $M$ UEs associated with a particular 5G AP, for example, the $n$th AP. Suppose at a given time slot $m$ ($m=1,\cdots, M$), this AP is transmitting to only one of the UEs via one beam in a round robin manner, which could be any one of the beams from $\mathbf{w}_{n1}, \cdots, \mathbf{w}_{nM}$, indexed as beam $1, \cdots, M$. During one scheduling cycle, the indices of the transmitted beams therefore form a beam sequence vector with a length $M$, denoted as $\mathbf{b}_n(t) = [b_{n1}(t), \cdots, b_{nM}(t)]^T$, where $b_{nm}(t)\in [1,\cdots, M]$. 


It is known that there are $\prod_{m=1}^Mm = M!$ permutations of such beam sequences, whereas there may exist only one optimal sequence given particular network criterion. This paper therefore aims at finding an optimal $\mathbf{b}_n(t)$ for the $n$th 5G AP, such that certain utility function is optimized in every scheduling cycle. In particular, we consider using the spectral efficiency as the utility function and aim to find an optimal beam sequence that maximizes the spectral efficiency. 

\section{Problem Formulation}
Let $\mathcal{B}$ denote the set that contains all $M!$ possible beam sequences. The mathematical description of the problem is given by
\begin{equation}
\hat{\mathbf{b}}_n = \argmax_{\mathbf{b}_n\in \mathcal{B}}U(\mathbf{b}_n)
\label{opt_problem}
\end{equation}
where $U(\mathbf{b}_n)$ is a utility function obtained when the sequence $\mathbf{b}_n$ is chosen as the beam sequence within one scheduling cycle with duration of $M$ time slot. When spectral efficiency is considered, the utility function for the entire scheduling cycle is given by
\begin{equation}
U(\mathbf{b}_n) = \frac{1}{M}\sum_{m=1}^MU(b_{nm}) 
\end{equation}
where $U(b_{nm})$ is the utility function for the $m$th user. We then consider the problem of finding the optimal $\hat{\mathbf{b}}_n$ such that the average spectral efficiency is maximized. 

\subsection{Derivation of Spectral Efficiency}
We now show the derivation of spectral efficiency $U(b_{nm})$. Suppose at a given time slot, the $m$th UE is scheduled and the 5G AP is transmitting via beam $\mathbf{w}_{nm}$. The spectral efficiency for the given time slot can be expressed as \cite{FM15}
\begin{equation}
U(b_{nm}) = \log_2\left(1+\frac{P_r(n,m)}{I(m)+N(m)}\right)
\label{untifunc}
\end{equation}
where $P_r(n, m)$ is the received signal power at the scheduled UE $m$, and $I(m)$ and $N(m)$ are the interference and noise term, respectively.

The received signal power is given by
\begin{equation}
P_r(n,m)(dB) = P_{n} + G_{n}(m) - PL(d) 
\end{equation}
where $P_{n}$ and $G_{n}(m)$ are the transmission power and beamforming gain at the $n$th 5G AP. In this paper we consider a constant transmission power, given by $P_{n} = \frac{P_{total}}{B}$, where $B$ is the bandwidth.

In addition, the beamforming gain at the $n$th 5G AP $G_{n}(m)$ is calculated as
\begin{equation}
G_{n}(m) = \left|\mathbf{w}^H_{nm}\mathbf{h}_{nm}\right|^2
\end{equation}
where $\mathbf{h}_{nm}$ is the channel between the $n$th base station to the scheduled UE given in \cite{FM15} as
\begin{equation}\label{hnm}
\mathbf{h}_{nm} = \sqrt{\frac{N}{L}}\sum_{l=1}^L\alpha_l\mathbf{a}_{UE}\left(\gamma_{l}^{UE}\right)\mathbf{a}^*_{AP}\left(\gamma_{l}^{AP}\right).
\end{equation}
In (\ref{hnm}), $\alpha_l$ is the complex gain of the $l$th  path, $\gamma_{l}^{UE}$ and $\gamma_{l}^{AP} \in [0, 2\pi]$ are the uniformly distributed random variables representing the angles of arrival and departure, respectively, and $\mathbf{a}_{UE}$ and $\mathbf{a}_{AP}$ are the antenna array responses at the UEs and 5G APs, respectively. Assuming uniform linear arrays, $\mathbf{a}_{AP}$ can be written as

\begin{equation}
\mathbf{a}_{AP} = \frac{1}{\sqrt{N_{AP}}}\left[1, \cdots, e^{j(N_{AP}-1)\frac{2\pi}{\lambda}D\sin(\gamma_l^{AP})}\right]^T\nonumber.
\end{equation}
For single antenna UE, we have 
\begin{equation}
\mathbf{a}_{UE} = 1\nonumber.
\end{equation}

Lastly, $PL(d)$ is the path loss component between the $n$th 5G AP and the $m$th user, which is a function of the distance $d$ between two nodes. We now look at the interference term given in (5), which is given by
\begin{equation}
I(m) = \sum_{n'=1\atop n'\neq n}^{N}P_r(n',m)
\end{equation} 
where $P_r(n',m)$ is calculated in the same manner as $P_r(n,m)$. The noise term $N(m)$ in (5) is simply white Gaussian noise, given by
\begin{equation}
N(m) = K_BTB
\end{equation}
where $K_B$ is the Boltzmann constant and $T$ is the noise temperature. 

Having obtained the spectral efficiency, the optimization problem given in (\ref{opt_problem}) can then be solved and the optimal beam sequence can be found. One could perform an exhaustive search in the finite set of possible beam sequences, known to yield a high computational complexity. In the following section, we propose a novel distributed learning algorithm to solve the optimization problem, which is shown to yield comparable performance than that using exhaustive search, while achieving a much reduced complexity.

\section{Beam Scheduling Algorithms}

We first present a distributed greedy scheduling mechanism, followed by a detailed description of the distributed learning algorithm. 

\subsection{Distributed Greedy Scheduling}
In the distributed greedy scheduling algorithm, at the beginning of each scheduling cycle, $\mathbf{b}_n$ is randomly chosen from the $M!$ possible permutation sequences for each 5G AP. A block-coordinate optimization algorithm is then applied to maximize the individual utility function of each 5G AP sequentially \cite{MHC15}. Different from exhaustive search, where a global optimization is reached and maximum spectral efficiency is achieved for all BSs, the distributed greedy scheduling mechanism maximizes the utility function with respect to $\mathbf{b}_n$ while keeping other $\mathbf{b}_i (i\neq n)$ unchanged. In other words, the $n$th 5G AP computes the utility functions for all possible permutations of $\mathbf{b}_n$, and then greedily selects the sequence that yields the maximum utility value, i.e., spectrum efficiency, for itself, assuming the first $(n-1)$ 5G APs are using the optimal sequences obtained in the previous selection process. The process continues until it reaches the last 5G AP, which completes one iteration of greedy selection. The same iteration will be repeated $N_{DG}$ times until a scheduling decision is made.

\subsection{Distributed Learning Scheduling}
In this section, we propose a distributed learning algorithm for beam scheduling. In the proposed learning algorithm, we allocate each sequence a probability at the beginning of each scheduling cycle and then update the probability and utility functions of the sequences iteratively. The optimal beam sequence is then selected at the end of the learning procedure according to such a probability. Such a learning algorithm is detailed as follows. 
 
Suppose the $k$th ($k\in [1,\cdots, M!]$) beam sequence is selected for 5G AP $n$ at iteration $t$, which we denote as $\mathbf{b}_{nk}^{(t)}\in \mathcal{B}$. At the beginning, i.e., $t=1$, each sequence is assigned with the same probability $p(U(\mathbf{b}_{nk}^{(1)}))=\frac{1}{M!}$, and one sequence $\mathbf{b}_{nk}^{(1)}\in \mathcal{B}$ is randomly selected for the $n$th 5G AP according to this probability. The utility functions are then calculated for each 5G AP. At the end of the $t$th iteration, the probability $p(U(\mathbf{b}_{ni}^{(t+1)}))$ ($1 \leqslant i \leqslant M!$) is updated for the $n$th 5G AP according to \cite{YR08} as
\begin{equation}
p(U(\mathbf{b}_{ni}^{(t+1)})) = p(U(\mathbf{b}_{ni}^{(t)}))-w\frac{U(\mathbf{b}_{ni}^{(t)})}{U^{max}(t)}p(U(\mathbf{b}_{ni}^{(t)}))
\end{equation}
subject to $\sum_{i=1}^{M!}p(U(\mathbf{b}_{ni}^{(t+1)}))=1$, where $i \neq k$, $w$ is a weighting factor, and $U^{max}(t)$ is the maximum utility function obtained up to iteration $t$, given by
\begin{equation}
U^{max}(t) = \max\{U(\mathbf{b}_n^{(1)}), \cdots, U(\mathbf{b}_n^{(t)})\}.
\end{equation}
For $i=k$, $p(U(\mathbf{b}_{nk}^{(t+1)}))$ is updated as
\begin{equation}
p(U(\mathbf{b}_{nk}^{(t+1)})) = p(U(\mathbf{b}_{nk}^{(t)}))+w\frac{U(\mathbf{b}_{nk}^{(t)})}{U^{max}(t)}P_n^{sum}
\end{equation}
where
\begin{equation}
P_n^{sum} = \sum_{i=1,i \neq k}^{M!}p(U(\mathbf{b}_{ni}^{(t)}))
\end{equation}

The similar learning procedure is applied to the next 5G AP until it reaches the last one and then the $(t+1)$th iteration starts. Such a learning process continues until the maximum number of iteration $T$ is hit and then the training phase stops. The final probabilities used to choose the optimal sequence for the $n$th 5G AP among all permutations is given by
\begin{equation}
\hat{k}_n(M) = \argmax_{k\in \{1,\cdots,M!\}}\{p(U(\mathbf{b}_{n}^{(1)})), \cdots, p(U(\mathbf{b}_{n}^{(M!)}))\}. 
\end{equation}   

\begin{figure} [t]
\centering
\includegraphics[scale=0.27]{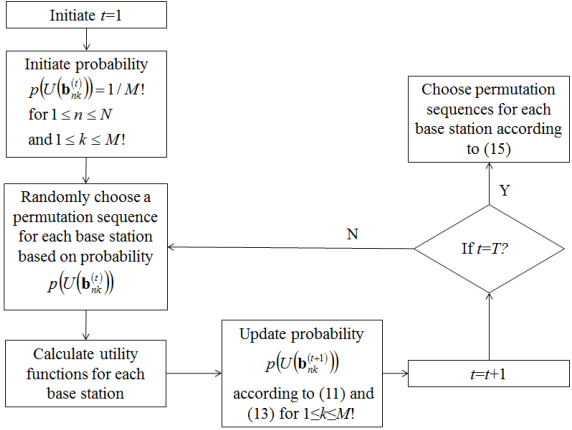}
\caption{Flow chart of the distributed learning scheduling algorithm}
\label{figflowchart}
\end{figure}

A flow chart of the distributed learning scheduling algorithm is given in Fig. \ref{figflowchart}. 

\subsection{Complexity and Signaling Overhead Analysis}


It is known that for exhaustive search, the statistical utility functions need to be computed for all APs and all possible beam sequences, yielding a complexity of $O(\left(M!\right)^N)$, which becomes prohibitive especially with a large number of APs. For the distributed greedy scheduling algorithm, for $N$ APs and $N_{DG}$ iterations, we need to calculate $N_{DG}NM!$ utility functions in total, having a computational complexity of $O(N_{DG}NM!)$. The proposed distributed learning algorithm, on the contrary, computes only one utility function for each 5G AP at a given iteration, yielding a complexity of $O(N_{LE}N)$, which is much smaller than the exhaustive search as well as the distributed greedy scheduling. In addition, as illustrated in the next section, the number of iterations required by the proposed leaning algorithm is also less than that of the greedy ones, i.e.,  $N_{LE} < N_{DG}$, leading to even less calculations.

In terms of signaling overhead, exhaustive search requires global utility function information to be exchanged among all 5G APs, whilst the signaling overhead of the proposed distributed learning scheduling algorithm is similar to the distributed greedy scheduling algorithm since there is no need for exchanging utility function globally. 

\begin{table}
\renewcommand{\arraystretch} {1.3}
\caption{Main System Parameters}
\label{table1}
\centering
\begin{tabular}{c c}
\hline\hline
Parameter & Value \\ 
\hline
Carrier frequency	& 60 GHz\\
Total bandwidth	    & 500MHz  \\
Base station Tx power 	& 30dBm\\
Inter-cell	distance & 200 or 400m\\
Number of base station antenna	& 8\\
Beam codebook size of base stations	& 16\\
Number of base stations	& 2 or 10\\
Number of UEs per base station	& 3 or 5\\
Number of UE antennas	& 1\\
Number of scatters 	& 3\\
Noise temperature    & 300 K\\
\hline
\end{tabular}
\end{table}

\section{Simulations}
In this section, we present simulation results obtained based on the scheduling algorithms proposed in the previous section. We assume a total transmission power of $30$ dBm and a total bandwidth of $500$ MHz and the 5G APs distribute the power uniformly over the entire bandwidth. The pathloss model used here is given in \cite{TSR13}.
The noise temperature $T$ is taken as the room temperature of $300K$. The detailed system parameters are presented in Table \ref{table1}.

\begin{figure} [t]
\centering
\includegraphics[scale=0.35]{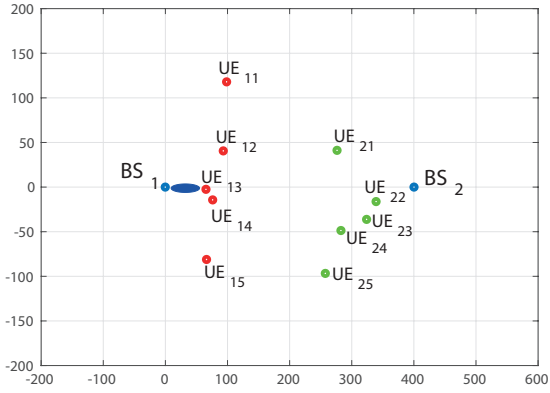}
\caption{Deployment of 2 5G APs (5 UEs per AP) }
\label{dep1}
\end{figure}



\begin{figure} [t]
\centering
\includegraphics[scale=0.35]{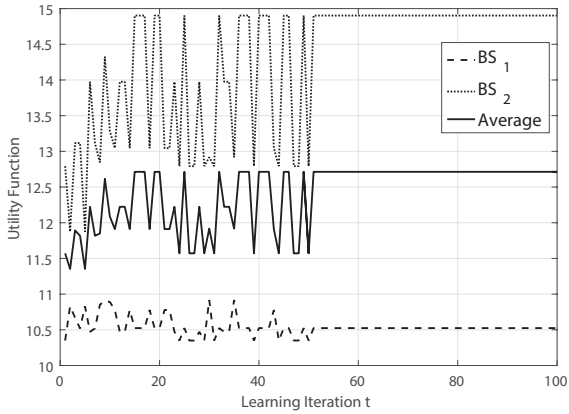}
\caption{Convergence behavior of the distributed learning scheduling algorithm}
\label{conv1}
\end{figure}

Fig. \ref{dep1} shows an example of a deployment scenario with 2 5G APs, each covering 5 UEs. In the figure, UE$_{ij}$ denotes the $j$th UE associated with the $i$th 5G AP. It can be seen that if UE$_{13}$ and UE$_{22}$ are scheduled at the same time, the transmission beam of AP$_{1}$ to UE$_{13}$ will cause interference to UE$_{22}$. Fig. \ref{conv1} illustrates the fluctuation of the utility functions of two APs obtained during the entire learning procedure with weighting factor  $w=0.15$. As illustrated, the utility functions rapidly converge to the optimal values in less than 55 learning iterations. The average utility function of two also converges to the maximum at the same pace.

%

\begin{figure} [t]
\centering
\includegraphics[scale=0.35]{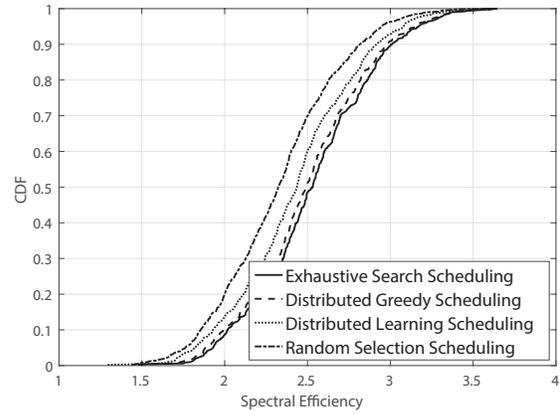}
\caption{CDF of spectrum efficiency (inter-cell distance = 400m)}
\label{CDF1}
\end{figure}

The cumulative distribution functions (CDFs) of utility functions for different scheduling algorithms are illustrated in Fig. \ref{CDF1}. For comparison purpose, we present the results obtained by using random selection of permutation sequences. As expected, the exhaustive search algorithm achieves the maximum overall utility function and the performance of the distributed greedy scheduling algorithm is slightly worse than the exhaustive one, which also serves as a performance boundary for all distributed scheduling algorithms. The performance of the proposed distributed learning algorithm is very close to the greed one but with significantly reduced complexity and signaling overhead as aforementioned. When the cell size is changed from 400m to 200m, the performance gap between the distributed greedy scheduling algorithm and the learning algorithm becomes even smaller.

%
%

\begin{figure} [t]
\centering
\includegraphics[scale=0.35]{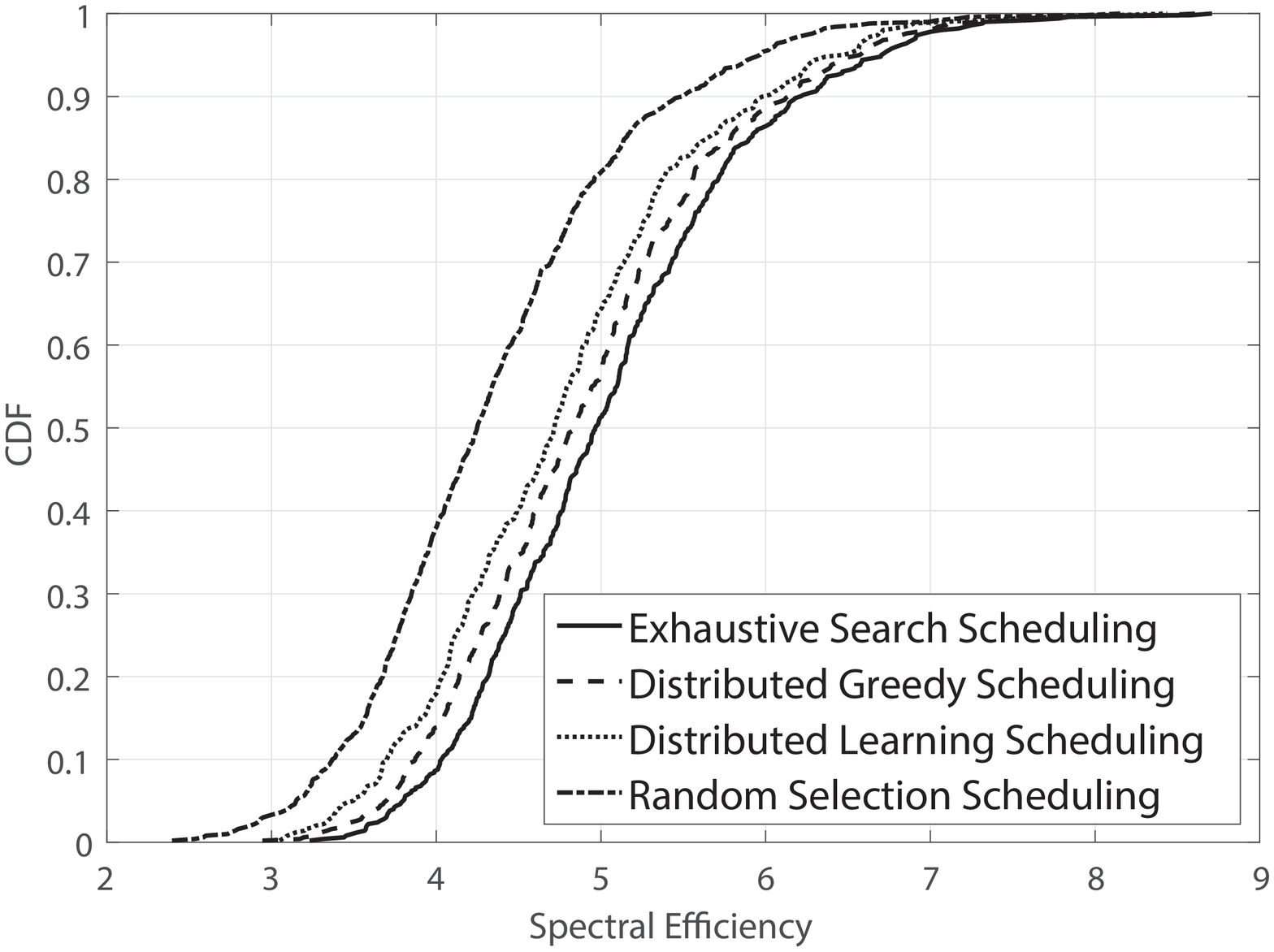}
\caption{CDF of spectrum efficiency (inter-cell distance = 200m)}
\label{CDF2}
\end{figure}

\begin{figure} [t]
\centering
\includegraphics[scale=0.35]{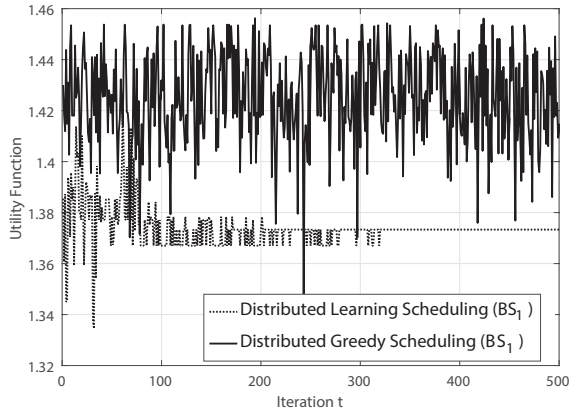}
\caption{Convergence behavior of two scheduling schemes}
\label{conv2}
\end{figure}

\begin{figure} [t]
\centering
\includegraphics[scale=0.35]{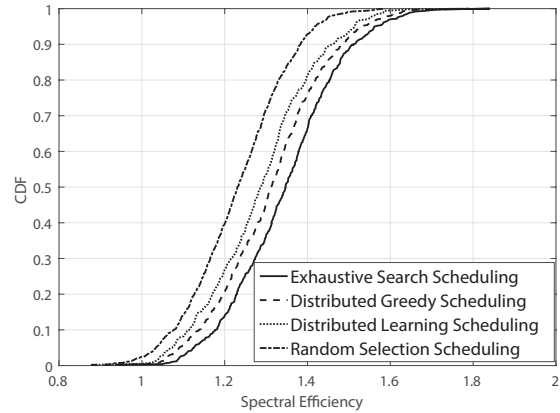}
\caption{CDF of spectrum efficiency (inter-cell distance = 400m)}
\label{CDF3}
\end{figure}

Then we increase the number of deployed 5G APs to 10, each covering 3 UEs, and clearly such deployment will result in a higher probability for a UE to receive interferences from other 5G APs. In Fig. \ref{conv2}, the convergence speed of the distributed greedy algorithm and the proposed learning algorithm is compared. Even though the greedy algorithm achieves a higher utility value, the convergence speed of the proposed learning algorithm is much higher (more than 150 iterations less), therefore leading to further reduced complexity. The CDFs of the utility function are illustrated in Fig. \ref{CDF3}. The proposed learning algorithm outperforms the random one and is close to the greedy one.

\section{Conclusion and Future Works}
In this paper, we propose a novel multi-RAT coexistence mechanism where neighboring 5G and WiGig APs, each serving their own associated UEs, schedule their beams in a distributed manner such that their own utility function, e.g., spectral efficiency, is maximized. The proposed distributed algorithm yields a comparable spectral efficiency for the entire networks as that using exhaustive search, which requires centralized coordination among multi-RAT networks with much higher algorithmic complexity. Our future work will focus on game theoretical analysis of our proposed algorithm with respect to fairness and its  convergence properties.

\section*{Acknowledgements}
The authors would like to thank Francesco Guidolin for his support and assistance with the simulations. The research leading to these results received funding from the European Commission H2020 programme under grant agreement n°671650 (5G PPP mmMAGIC project).
 
\end{document}